# Shakedown: compiler-based moving target protection for Return Oriented Programing attacks on an industrial IoT device

Fady Copty[*,1], Francisco Hernandez[2], Dov Murik[1], Olmo Rayón[2]

[1] IBM Research – Haifa, Israel
[2] Worldsensing, Spain
[*] `fadyc@il.ibm.com`

**Abstract.** Cybercriminals use Return Oriented Programming techniques to attack systems and IoT devices. While defenses have been developed, not all of them are applicable to constrained devices. We present Shakedown, which is a compile-time randomizing build tool which creates several versions of the binary, each with a distinct memory layout. An attack developed against one device will not work on another device which has a different memory layout. We tested Shakedown on an industrial IoT device and shown that its normal functionality remained intact while an exploit was blocked.

## 1    Introduction

The Industrial Internet of Things (IIoT) relies on the network of physical devices and the data exchange between them, which results in economic benefits by increasing the automatization and control of labor-intensive processes. IIoT meets the demands of many industry verticals and it arises as the natural evolution and merging of information technologies, wireless connectivity and industrial control techniques. However, the level of maturity of IIoT is still in its infancy that together with the rapid and, why not to say, disorderly growth has led to potential issues with a direct impact on citizens' safety, security and privacy.

When approaching the challenge of IIoT security, it is important to note that this goes beyond the standard concepts of cybersecurity and involves many other disciplines. In particular, the physical perspective is often put aside in the efforts to ensure the integrity, confidentiality and availability of the information. Nevertheless, the physical compromise of the endpoints can seriously harm the data flow and as result jeopardize the functioning of the whole system.

The adoption of security measures in IIoT ecosystems is bound to intrinsic technological restrictions such as the limited computational power of the devices, as well as energy consumption constraints due to the typical batteries-operated designs. In this scenario, most of the organizations working with IIoT technologies prioritize the sys-tem functionalities (i.e. lifespan, data sampling rate) in front of the adoption of secure elements such as cryptography, message signatures and keys interchange. It goes without saying that this situation is not optimal and an accurate risk analysis to detect and handle the potential vulnerabilities, trying to reach an equilibrium with the business requirements, should be an accepted practice. This would definitely improve the maturity level of IIoT technologies and their promotion in critical infrastructures.

Almost 90 percent of exploit-based software attacks use hostile Return Oriented Programing (ROP) techniques in their attack chain [10]. Return Oriented Programing is an attack that hijacks the control flow of a program to allow remote code execution by an attacker. Over the years, much work has been done to break this control flow hijack phase of the attack chain and prevent remote code execution of malicious code. So far, the available techniques have had limited success, and have introduced a significant performance burden on the protected devices.

Return Oriented Programing (ROP) is the most common control flow hijack exploit technique that allows an attacker to take control of a program flow. This is done by smashing the call stack and redirecting it to execute attacker instruction sequences. The attacker caries out this exploit by borrowing gadgets, or small pieces of code, from the hijacked program and redirects the control flow to execute these gadgets. The attacker malicious code is compiled of a sequence of these gadgets.

Compiling a ROP attack requires an offline phase where attackers read the target program binary, find the desired gadgets in the target program, and create a list of entry points to execute these gadgets. Each one of these gadgets ends with a return instruction and the attacker hijacks the control flow by controlling the return address of these gadgets, hence the name return-oriented-programing.

Cybercriminals use ROP because it can easily bypass data execution prevention (DEP), a technology implemented in hardware and software to prevent execution of data sections. One of the many techniques developed to prevent ROP attacks is address space layout randomization (ASLR), a process implemented in almost all operating systems. ASLR randomizes the address bases of the sections, forcing the attacker to guess the location of the gadgets. Attackers building a ROP attack assume that the gadgets they need are in absolute addresses or shifted by constant bytes. If they can detect the shift of one gadget, they can detect the shifts of all gadgets.

This is where our novel moving target solution comes in. Given a source code of a program and a randomization seed we compile the source code with randomization of: the order of the functions, the alignment of functions, and the linking order of the objects. For every deployment of the program on an IoT device we compile a new unique copy of the program.

Our solution challenges the attacker to guess all address locations of the gadgets. Unlike in ASLR technology, finding a gadget in one block does not reveal the addresses of the other gadgets. This results in a new situation where every IoT device is created in a unique manner and renders the attacker guessing of gadget addresses impossible.

The Shakedown solution can block memory corruption zero-day attacks. Beyond that, it can detect attack attempts at almost no performance cost and without having to install new online components.

## 2 Shakedown implementation

We will now describe the implementation details of Shakedown, our anti-ROP solution [5]. Shakedown is a build toolchain (compiler and linker) which randomizes the location of pieces of code inside the resulting binary executable. The randomization is based on an externally-supplied seed value; supplying different seed values will cause the toolchain to generate distinct binary executables from the same source.

Shakedown employs three randomization techniques: function order randomization, function address alignment randomization, and object linking order randomization. The first two are implemented as compiler passes inside the LLVM [7] compiler framework, whereas the latter is implemented as a wrapper to the system linker. We will now describe the randomization techniques in detail.

### 2.1 Randomization techniques

In order to further explain the operation of the different randomization techniques, we compile the following test program with the different randomizers active:

```
#include <stdio.h>

void func_a() {
  printf("inside func_a\n");
}

void func_b() {
  printf("inside func_b\n");
}

void func_c() {
  printf("inside func_c\n");
  func_b();
  func_a();
}

int main() {
  func_a();
  func_b();
  func_c();
  return 0;
}
```

**Function order randomization.** This randomization technique is implemented as an LLVM compiler pass executed before the machine-specific backend code generation starts. It creates a random permutation of the functions in the module and passes the shuffled list of functions to the next compiler pass. This causes the generated binary code to include the machine code for the module's functions in a different order, without harming functionality.

**Table 1** shows the addresses of the functions in the resulting binary file. We see that the first function always starts at 0x00010414, but the order of the functions is different in each of the builds, and therefore the code that exists in any given address in memory is unpredictable.

**Table 1.** Addresses of functions using function order randomization

| Original build | Shakedown build seed=42 | Shakedown build seed=97 |
| --- | --- | --- |
| 00010414 func_a | 00010414 main | 00010414 main |
| 00010438 func_b | 00010444 func_c | 00010444 func_c |
| 0001045c func_c | 00010470 func_a | 00010470 func_b |
| 00010488 main | 00010494 func_b | 00010494 func_a |

**Function address alignment randomization.** The LLVM framework allows the user to define an alignment for each function; for example, an alignment of 16 instructs the compiler to place the function's start at an address divisible by 16. We implemented an LLVM compiler pass which randomly enlarges the alignment definitions of module's functions (note that only powers of two are valid alignment values), and thereby creates a space between functions which is different for each Shakedown build. Note that by only enlarging the alignment we keep the user's request for alignment (for example, a user alignment request of 16 is still met if Shakedown enlarges it to 32 or 64).

We enabled this compiler pass and examined the addresses of the functions in the resulting binary file, as seen in **Table 2**. We see that with this randomization technique, the functions remain in the same order as in the original binary, but their exact location is shifted due to the different alignments inserted by this technique.

**Table 2.** Addresses of functions using function address alignment randomization

| Original build | Shakedown build seed=42 | Shakedown build seed=97 |
| --- | --- | --- |
| 00010414 func_a | 00010480 func_a | 00010480 func_a |
| 00010438 func_b | 00010500 func_b | 00010500 func_b |
| 0001045c func_c | 00010524 func_c | 00010540 func_c |
| 00010488 main | 00010560 main | 00010570 main |

**Object linking order randomization.** The two randomization techniques we explained so far only shuffle code inside one module; but the order of the modules (object files) inside the resulting binary stays unchanged. In order to further shuffle the resulting binary, we added a wrapper to the linker that will randomize the order of the object files supplied to the linker and then execute the original linker with the shuffled object files.

In order to show the effect of this technique, we separated the original source code into 4 source files with one function in each. We then compiled each source file to its own object file, and then linked the 4 object files into one executable binary file. We then examined the addresses of the functions in the resulting binary files, as seen in

**Table 3**. Shakedown randomizes the linking order of the 4 object files, resulting in different locations for the different objects, and therefore, different locations for the functions implemented in those objects.

**Table 3.** Addresses of functions using object linking order randomization

| Original build | Shakedown build seed=42 | Shakedown build seed=97 |
|---|---|---|
| 00010414 func_a | 00010378 func_b | 00010378 main |
| 00010438 func_b | 00010464 func_a | 00010510 func_a |
| 0001045c func_c | 000104ac func_c | 00010534 func_c |
| 00010488 main | 000104d8 main | 00010560 func_b |

### 2.2 Deployment options

All the different randomization techniques use random values generated by a standard pseudo random number generator (PRNG) seeded by an integer value received from an environment variable. This way, the developer may create N distinct copies of the target binary executable by running the toolchain N times with distinct seed values.

For additional security, developers may also use Shakedown to compile libc (or its small-footprint equivalents such as uClibc-ng [11] or musl [8] and statically link it with the target program. This prevents a ROP attack which relies on gadgets inside libc's code.

In an IoT setting, ideally each device is loaded with a distinct executable binary. This way, an attacker learning memory addresses of functions or ROP gadgets from one device will gain no information regarding equivalent memory address on another device, because the memory layout of the target program is different than the one the attacker examined.

## 3    Experiment

The piece of hardware used in the tests described in the article is a customized Kerlink gateway customized which fulfils the requirements from Worldsensing's Industrial IoT solution (Loadsensing). The gateway is ready for working outdoors and it offers 8-band capacity, onboard GPS, and the following communication elements: Ethernet or cellular network to connect with the Internet and LoRa for the endpoints (sensors and dataloggers). LoRa is a long range and low power protocol that enables communications between remote IoT devices by using a unique modulation format.

The role of the gateway in the Loadsensing architecture is to link the sensors with the cloud. As described before, the data generated in the endpoints is received as LoRa packages in the gateway, where it is processed to optimize the amount of information transmitted to the cloud. To that end, an edge-computing approach is followed. The gateway is therefore the first element of the architecture with enough computational power to be considered critical. Its malfunction jeopardizes the quality of service of Loadsensing, causing a lack of control over the distributed sensors.

To show the effectiveness of Shakedown, we needed a memory-related vulnerability in the gateway. We artificially added a buffer overflow vulnerability in which data from an incoming packet is copied into a local buffer without size checking. To simplify the test, we also added a secret target function which the attacker wants to call (in a real-life scenario this can be an admin-only function, or simply a sequence of ROP gadgets to perform the attacker's shellcode).

In our control case, we compiled the gateway's source code with clang targeting the ARM926E-JS CPU and installed the binary in the gateway. We crafted the attack packet which includes the long buffer with the address of the secret target function in it. When we sent that malicious packet to the gateway, the packet's data was copied into the local buffer, the return address on the stack was overwritten with the address of the secret target function, and when the function returned the secret function was called, which means the attack was successful.

We then compiled the same source code with Shakedown and installed the shuffled binary inside the gateway. When we sent normal packets, the gateway behaved exactly as the original gateway; when we sent the attack packet, the gateway's program crashed (and immediately restarted) but the secret target function was not called. The attack was not successful, because the attack packet was developed to successfully attack one gateway with its own memory layout; the same packet could not be used to attack another gateway which has a different memory layout due to the Shakedown randomization.

## 4     Related work

A lot of research has been done in this area of exploit mitigation using moving target techniques. In general, these techniques can be categorized into two main categories: binary-based approaches and compiler-based approaches.

Binary-based approaches operate directly on the binary. Address Space Layout Permutation (ASLP) was introduced by [5]. This technique performs function permutation of the binary and allows re-randomization on each run of the program. A different technique was introduced in [9] where the authors use in-place code randomization to within the same block. Another technique is instruction location randomization (ILR) [4], where every instruction location is randomized, and a new execution scheme is introduced. However, the main drawback of the binary-based approach is that it requires disassembly of the binary to identify code segments and relocation information. Disassembly is a hard problem that often relies on heuristics, and the disassembled code is usually not reliable for this kind of randomization where the randomized program must maintain the same functionality of the original.

Compiler-based approaches operate on source code or on compiled objects. Authors of [3] explore the feasibility questions of software diversity in the mobile market, and contributes to the argument that a paradigm shift in software distribution is feasible for mobile applications through a diversification engine operating at centralized software distribution center (e.g. App Store). In this paper we will present a de-vice diversity paradigm, where each device's software is randomized from source code, we will also report experimental deployment made on an industrial IoT device.

Finally, in [1] authors introduce a self-randomizing transformation, where they compile a single binary from source code while adding a self-randomization capability to this binary. This results in the ability to maintain the single binary copy distribution model, while enabling randomization at run time or load-time. One example using a similar technique is a linker-based solution Kernel Address Randomized Link (KARL) that was introduced by OpenBSD [2]. KARL re-links the kernel on every boot to randomize the order of object files inside the kernel binary. The main drawback of this technique is the overhead it introduces to the IoT device.

## 5   Conclusions

This paper presents a tool for preventing memory access attacks including ROP using various compiler-time and link-time randomizations. By generating many versions of the binary which are functionally identical but have completely distinct memory layouts, attackers cannot develop a ROP attack by examining one binary. Such a scheme is especially useful for IoT devices which might not employ other defenses such as ASLR. The tool was tested successfully on an industrial IoT device: Worldsensing's Loadsensing gateway.

In the future, this work can be extended to add more randomization techniques to further increase the variability of generated binaries. Moreover, illegal memory access traps may be added automatically inside the code to detect an exploit which attempted to execute code gadgets outside the normal program's code; such a trap can then report the attack to a security events auditing system which can cross reference it with other activity in the system.

Deploying randomized binaries to different devices requires special kind of management that should be addressed, especially with respect to investigating field bug reports and providing system and application updates. The fact that each device is different creates a challenge for existing processes, which require adaptation for this new moving target scheme.

## 6   Acknowledgements

This project has received funding from the European Union's Horizon 2020 research and innovation programme under grant agreement No 740787 (SMESEC). We would like to thank Ayman Jarrous for developing an earlier version of this project and for the fruitful discussions.